\documentstyle[erice97,12pt]{article}

\input epsf.tex
\def\ie{{\it i.e.\/}}

\def\etal{{\it et.al.\/}}

\def\roughlyup#1{\mathrel{\raise.3ex\hbox{$\sim$\kern-.75em
\lower1ex\hbox{$#1$}}}}
\def\roughlydown#1{\mathrel{\raise.3ex\hbox{$#1$\kern-.75em
\lower1ex\hbox{$\sim$}}}}

\def\lsim{\roughlydown<}

\def\Avg#1{\langle #1 \rangle}

\def\eqa{\begin{eqnarray}}
\def\eeqa{\end{eqnarray}}
\def\eq{\begin{equation}}
\def\eeq{\end{equation}}

\def\Tr{{\rm Tr}\,}

\def\Sca{{\cal A}}
\def\Scc{{\cal C}}
\def\Scd{\sigma^2}
\def\Sco{{\cal O}}
\def\ssv{{\sss V}}

\def\sss{\scriptscriptstyle}

\def\GF{G_{\sss F}}

\def\dv{\delta V}
\def\dne{\delta n_e}

\def\MSW{{\sss MSW}}
\def\VAC{{\rm vac}}
\def\osc{{\rm osc}}

\begin{document}
\title{Solar Fluctuations and the MSW Effect}
\author{C.P. Burgess}
\address{Physics Department, McGill University, 3600 University Street,\\
Montr\'eal, Qu\'ebec, Canada, H3A 2T8.}

\maketitle

\abstracts{This talk summarizes the results of
recent calculations of how fluctuations
within the solar medium can influence 
resonant neutrino oscillations 
within the sun. Although initial calculations
pointed to helioseismic waves as possibly
producing detectable effects, recent more 
careful calculations show this not to be true. 
Those features of fluctuations which maximize 
their influence on neutrino propagation are
identified, and are likely to have
implications for supernovae and the early universe.}

\section{Introduction}

The results from solar-neutrino 
experiments \cite{snexps} now strongly suggest that
the experimentally-observed solar-neutrino spectrum differs 
from the predictions of astrophysical theory \cite{SNTheory}. 
The observed spectrum is consistent with a virtually complete 
elimination of the flux of intermediate-energy ${}^7$Be 
neutrinos, together with a partial suppression of the higher-energy 
${}^8$B neutrino flux. This kind of pattern is notoriously difficult
to obtain using astrophysical variations on the solar-model
theme (notwithstanding the recent proposal of significant 
${}^3$He convection \cite{He3conv}), suggesting instead
an interpretation based on nonstandard neutrino properties.
Perhaps the most successful such neutrino-based explanation
is the MSW mechanism \cite{msw}, which proposes the 
resonant conversion of neutrino flavours within the solar medium,
providing an excellent description of the experiments 
using theoretically-plausible neutrino parameters.

The emergence of the MSW mechanism as the leading explanation
for the solar-neutrino problem has motivated more detailed 
scrutiny of the approximations on which its predictions 
rest. For example, the original MSW analysis 
starts from a mean-field treatment of the background 
through which the neutrinos propagate. More recent
work has since investigated how corrections to the mean-field
picture might influence the neutrino survival probability
\cite{sawyer, noisy, noisytwo, 
cliffdenis, swiss, petercliffdenis}.
These studies indicate that there are astrophysical
situations for which fluctuations might significantly
influence neutrino propagation despite the extreme weakness
of neutrino interactions. The two different ways 
in which they might do so are:

\begin{enumerate}

\item
Fluctuations can scatter neutrinos away from the forward
direction, and so shorten the effective neutrino mean free
path a the medium. 

\item
For situations where the mean particle densities 
cause resonant oscillations amongst neutrino
flavours, fluctuations generically act to ruin the 
quality of the resonance. 

\end{enumerate}

Explorations of the practical implications of these 
effects are in their infancy. Naturally enough, the
first applications have been to solar neutrinos, for
which the most experimental information is 
available.\footnote{See, however, ref.~\cite{spinfl}
for a discussion of some fluctuation effects within
supernovae.} They have therefore focussed on the second of the 
alternatives listed above, since the density of the 
solar medium precludes seeing the consequences of direct 
neutrino scattering from fluctuations.

By contrast, MSW oscillations, if they occur, hold out
the most promising way of seeing the effects of solar 
fluctuations on neutrino propagation. This is because
even though the fluctuation effects are small, 
resonant oscillations make neutrinos more 
sensitive to fluctuations 
than they would otherwise be. Most
interestingly, independent preliminary calculations 
\cite{noisytwo, cliffdenis} 
indicated that reasonbly-sized
fluctuations in the solar interior might be 
detectable through the deterioration of the
MSW resonance which they induce. 

The next sections contain a summary of
these estimates as well as subsequent, more
precise, calculations \cite{petercliffdenis} of the
effects of known solar fluctuations on neutrino
propagation. In so doing we follow the presentation
of refs.\cite{cliffdenis} and \cite{petercliffdenis}, 
with which we are most familiar. \S2 summarizes
which features fluctuations must have in order to
appreciably affect neutrino oscillations. For
solar applications, helioseismic waves emerge
as the fluctuations with the best chance of 
spoiling the MSW effect.  
\S3 then briefly describes the results of detailed 
simulations of neutrino oscillations in the presence
of helioseismic waves, culminating with the 
conclusion that no known source of solar fluctuations
is large enough to have a detectable influence on
solar neutrinos.

\section{General Features}

Much is known about how particles propagate
through materials, and general techniques have 
been developed to describe this propagation. 
This section describes the adaptation of these 
ideas to neutrino physics, as developed in 
ref.~\cite{cliffdenis}. 

\subsection{The Origin of Fluctuations}

Since neutrinos interact so weakly with the
solar medium, they are negligibly scattered
by microscopic effects such as thermal 
fluctuations. As a result, the time evolution of 
the density matrix, $\rho$, describing the neutrino state
is given by the usual quantum-mechanical expression:
\begin{equation}
\label{Schreqn} 
\rho(t,n) = U(t,n) \, \rho(0) \, U^\dagger (t,n),
\end{equation}
where $t$ denotes time, $U(t,n)$ is the 
evolution operator, and $n$ generically denotes 
all of those features of the solar medium (like
the local electron density profile) on which 
neutrino evolution can depend parametrically.
For neutrino oscillation experiments we take
$\rho$ to be a matrix in neutrino-flavour space, 
and $U$ is the solution to the evolution equation
\begin{equation}
\label{Ueveqn}
{\partial U \over \partial t} = -i V(t,n) \, U
\end{equation}
where $V(t,n)$ is the interaction Hamiltonian 
which couples the neutrinos to each other and
to the solar medium. For example, for an 
ultrarelativistic neutrino having 
three-momentum $k$ and mass matrix $m$ propagating
through the sun, the dominant flavour-dependent 
interactions are $V =  V_\VAC + V_{\rm mat}(t,n)$,
with:
\begin{eqnarray}
\label{Vdefs}
V_\VAC &\approx& {m^\dagger m\over 2k}+\dots \cr
V_{\rm mat}(t,n) &\equiv& \sqrt{2}\,\GF g^e n_e(t),
\end{eqnarray}
where $g^e={\rm diag}(1,0)$ is a matrix describing
the charged-current coupling to the electron density
$n_e$. For simplicity we consider here only
the case of two active neutrino species.

For the present purposes fluctuations arise 
when the electron density, $n_e$, is
not constant in space or time, since then the 
solar properties as seen along the trajectories
of successive neutrinos can vary. The
response of a detector to many such
neutrinos is then described by an
average over the ensemble of densities
which are seen by the individual particles,
weighted by some probability
distribution, $p(n)$:
\begin{equation}
\label{obsavg}
\Avg{\Sco}(t) = \int dn \; p(n) \; \Tr \Bigl[ \Sco
\;  \rho(t,n) \Bigr] = \Tr \Bigl[ \Sco
\;  \Avg{\rho(t)} \Bigr] .
\end{equation}

The physics of the fluctuations is encoded in
the probability distribution, $p(n)$, which we
define in terms of the expansion of $n_e$ in
a complete set of basis functions, $\phi(t)$,
that are assumed to describe uncorrelated variables.
That is:
\begin{equation}
\label{basis}
n_e(t) = \Avg{n_e(t)} \left[1+\sum_j \Scc_j \phi_j(t) \right] ,
\end{equation}
where the coefficients $\Scc_j$, are assumed to be
uncorrelated random variables which are Gaussian distributed,
with vanishing mean: $\Avg{\Scc_j}=0$  and
$\Avg{\Scc_j\Scc_k}=\Scd_j\delta_{jk}$.

The most commonly-used choice of basis functions
\cite{noisy, noisytwo, cliffdenis} are
localized in space, corresponding to 
fluctuations which are uncorrelated from point to point.
Since this choice is {\it not} appropriate for helioseismic
waves, for these we instead follow refs.~\cite{cliffdenis,
petercliffdenis} and expand the electron density in terms
of a complete basis of helioseismic modes \cite{solwaves}.

The brute force approach is to directly solve these
equations for $U(t,n)$ and $\rho(t,n)$, to find the 
electron-neutrino survival probability for any one
neutrino after its transit through the sun, and to 
then numerically average the result over the appropriate
ensemble. Before describing the results so obtained,
we first pause to describe an analytic solution which
applies in the limit that the correlation lengths 
describing the fluctuations are much smaller than 
all other scales in the problem.  

\subsection{Short Correlation Lengths: The Master Equation}

It pays to recognize that the neutrino-oscillation 
problem in a random medium necessarily involves (at 
least) two important time scales. This pays because
the time-evolution greatly simplifies if one of these
scales should be much smaller than the other. 
All of the calculations of solar fluctuations,
apart from ref.~\cite{petercliffdenis}, are based on
this simplifying assumption. This section describes 
how this approximation works. 

One of the basic scales for neutrino oscillations
is the oscillation time, $\tau_\osc$, which we define to be
the time required for different neutrino flavours
to acquire significantly different phases as they
evolve through the medium. By definition, 
oscillation phenomena always involve evolution
over scales $t > \tau_\osc$, 
which cannot be computed perturbatively
in the interactions which distinguish between the
oscillating neutrino species.  

The randomness of the interaction between the
neutrino and its environment, described above, 
introduces the other important timescale. 
More precisely, denoting $\dv = V - \Avg{V}$, 
suppose that the correlation $\Avg{\dv(t) \, \dv(t')}$ 
is negligible whenever $|t - t'|$ is greater than 
some characteristic scale, $\tau_c$. In this case the
neutrino density matrix at time $t$, $\rho(t)$, 
retains a memory of the fluctuations through which
it has passed only over the time interval
$(t - \tau_c,t)$.

The simplification occurs if $\tau_c \ll \tau_\osc$.
In this case, the evolution from an initial 
state, $\rho(t)$, to a state at
a later time, $\rho(t + \Delta t)$, can be computed 
in perturbation theory: 
\eq
\label{DrDtdef}
\Delta\rho \equiv \rho(t + \Delta t) - \rho(t) \equiv {D\rho
\over Dt} \; \Delta t,
\eeq
so long as $\Delta t \ll \tau_\osc$. Here $D \rho/Dt$ 
{\it defines} a coarse-grained time derivative, which
may be computed as an explicit convolution of $\rho$ and
$V$ over the time interval $(t, t+\Delta t)$. 

On the other hand, if $\Delta t$ can also be
chosen to satisfy $\Delta t \gg \tau_c$, then 
$D \rho/Dt$, evaluated at time $t$, depends only 
on the initial value of $\rho(t)$, and
has no memory of how $\rho$ has evolved over
earlier times. This removes the complication of
having $D\rho/Dt$ involve a convolution over
the entire time interval, $(t, t+\Delta t)$.
For example, as applied to the neutrino density 
matrix in two-by-two flavour-space, the twin conditions
$\tau_c \ll \Delta t \ll \tau_\osc$ lead to:
\begin{equation}
\label{drhoflavour}
{D \rho \over Dt} =-i \Bigl[ V_\VAC + V_\MSW(t) ,\rho \Bigr]
-2 \, \GF^2 \Sca(t) \Bigl[ (g^e)^2\rho + \rho (g^e)^2 - 2 \, g^e
\rho g^e \Bigr] + {\cal O}(V^3).
\end{equation}
where $V_\MSW(t,n) = \sqrt{2}\,\GF g^e \Avg{n_e(t)}$
and the coefficient $\Sca(t)$ of the
second term is the correlation integral
\begin{equation}
\label{correlation}
\Sca(t) \equiv\int_{t'}^t  d\tau \; \Avg{\dne(t)\dne(\tau)}
\end{equation}
which represents fluctuation effects. Notice that the first term 
on the right-hand-side of eq.~(\ref{drhoflavour}) 
describes the usual MSW evolution of neutrino flavour.

Now comes the main point. Eqs.~(\ref{DrDtdef}) and 
(\ref{drhoflavour}) --- the {\it Master 
Equations} --- define a Markov-like process
which gives $\partial\rho(t)/\partial t \approx 
D\rho/Dt$ purely in terms of $\rho(t)$ and $V(t)$. 
This relation may be directly integrated 
to obtain $\rho(t')$ for {\it any} later $t'$. The
result holds even for $|t' - t| \gg \tau_\osc$, even
though $D\rho/Dt$ is only computed perturbatively 
in $V$.

For two-flavour neutrino oscillations this master
equation, including fluctuations, may be solved 
analytically subject to the same assumptions as
are typically used to solve the MSW oscillation
problem \cite{cliffdenis}. The resulting expression 
for the electron-neutrino
survival probability is a generalization of 
the well-known Parke formula \cite{Parkeref} to 
include the effects of fluctuations:
\begin{equation}
\label{parke}
P_e(t,t') = {1\over 2} + \left( {1\over 2} - P_J \right) \lambda
\cos 2\theta_m(t')\cos 2\theta_m(t) .
\end{equation}
Here $P_J = \exp\left[ -{\pi\over 2} 
\left({\sin^22\theta_\ssv \over
\cos 2\theta_\ssv} \right) \left( {\delta m^2 h 
\over 2k} \right) \right]$ is the `jump' probability, 
where $h$ is the scale height for $\Avg{n_e(t)}$, 
and $\delta m^2$ is the 
squared-mass difference between the two neutrino 
mass eigenstates in vacuum. $\theta_\ssv$ is the 
vacuum mixing angle, while $\theta_m(t)$
is the matter mixing angle evaluated at the 
position occupied by the neutrino at time $t$. 
The contributions of fluctuations are summarized
by the coefficient $\lambda$, which is given by
\begin{equation}
\label{lamb}
\lambda\equiv \exp\left[ -2 \, \GF^2 \int_{t'}^t  d\tau \;
\Sca(\tau) \sin^22 \theta_m(\tau) \right] .
\end{equation}
An important consequence of this equation
is its implication that the probability of $\nu_e$ 
survival depends almost exclusively on fluctuation 
properties at the position of the MSW
resonance, since $\sin^22\theta_m$ is typically sharply
peaked at this point.

Figure 1 plots the survival probability {\it vs}
neutrino energy 
which is predicted for two particular models of 
solar density fluctuations. Successive curves in this
plot represent fluctuations having different values
for their correlation length, $\ell$, and their 
relative amplitude, $\epsilon^2 \sim \Avg{\delta n_e
\delta n_e}/\Avg{n_e}^2$ at
the position where the MSW resonance occurs.

\vspace{0.45cm}

\centerline{\epsfxsize=9.5cm\epsfbox[45 430 550 750]{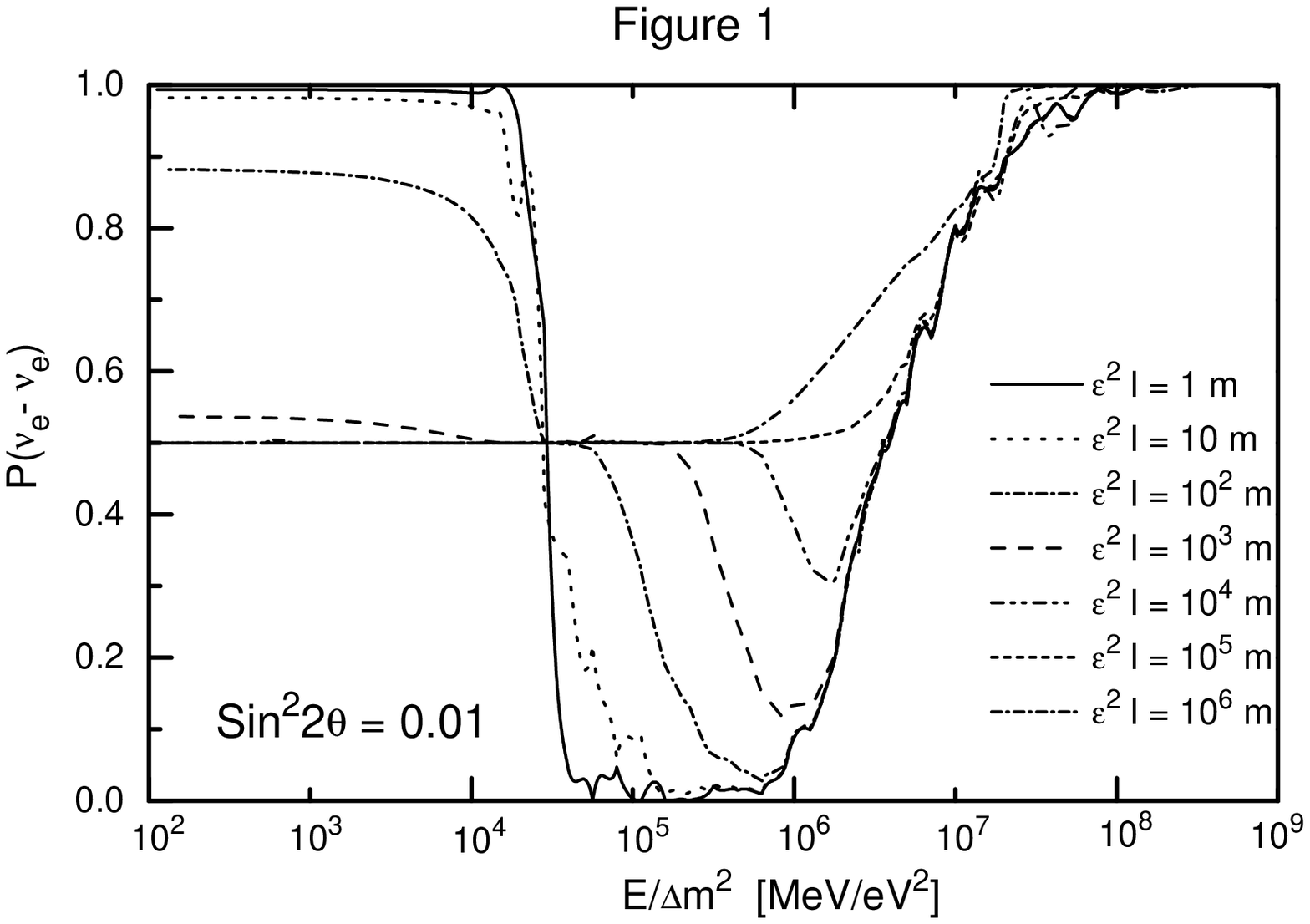}}

\begin{quote}
{\footnotesize  {\bf Figure 1:} {\sl The electron-neutrino survival rate
{\it vs} neutrino energy as 
predicted by the Master-equation method using a
particular model of solar fluctuations, and
representative solar-neutrino properties.
Each curve represents a fluctuation having a
different value for the one relevant combination
of amplitude, $\epsilon$, and correlation
length, $\ell$. See ref.~\cite{cliffdenis} for
details.}}
\end{quote}

\subsection{Long Correlation Lengths}

The generalized Parke formula provides an
excellent description of neutrino oscillations in
media having short-correlation-length fluctuations,
as has been verified by comparing it with direct 
numerical integrations \cite{cliffdenis}. 
Two important questions remain, however, which
the present section addresses. First, how
badly does it fail when $\tau_c$ is {\it not} small
enough to justify the Master-equation approach? 
Second, for realistic solar neutrinos what does 
the condition $\tau_c \ll \tau_\osc$
translate into for the correlation length
of a candidate fluctuation in the sun?

Anwering the second of these first, for solar
neutrinos the correlation time, $\tau_c$,
represents the time taken for a neutrino to
traverse the correlation length, $\ell$, of
a solar fluctuation. The numerical comparisons
and analytical estimates of ref.~\cite{petercliffdenis}
indicate that the Master 
equation may be expected to work so long as 
$\ell$ satisfies $\ell \lsim \ell_{\rm crit} =
1/(\epsilon\GF 
\Avg{n_e})$ (recall $\epsilon^2 \sim \Avg{\delta n_e
\delta n_e}/\Avg{n_e}^2$ is a measure of the
relative amplitude of the fluctuation at
the position where the MSW resonance occurs). 
Taking $\epsilon \sim 10 \%$ for the fluctuation
amplitude and $\ell_{\rm crit} \sim 300{\rm km}$,
corresponding to the electron density
at the resonance point, this condition becomes
$\ell \lsim 3,000 {\rm km}$. 

\vspace{0.5cm}

\centerline{\epsfxsize=9.5cm\epsfbox{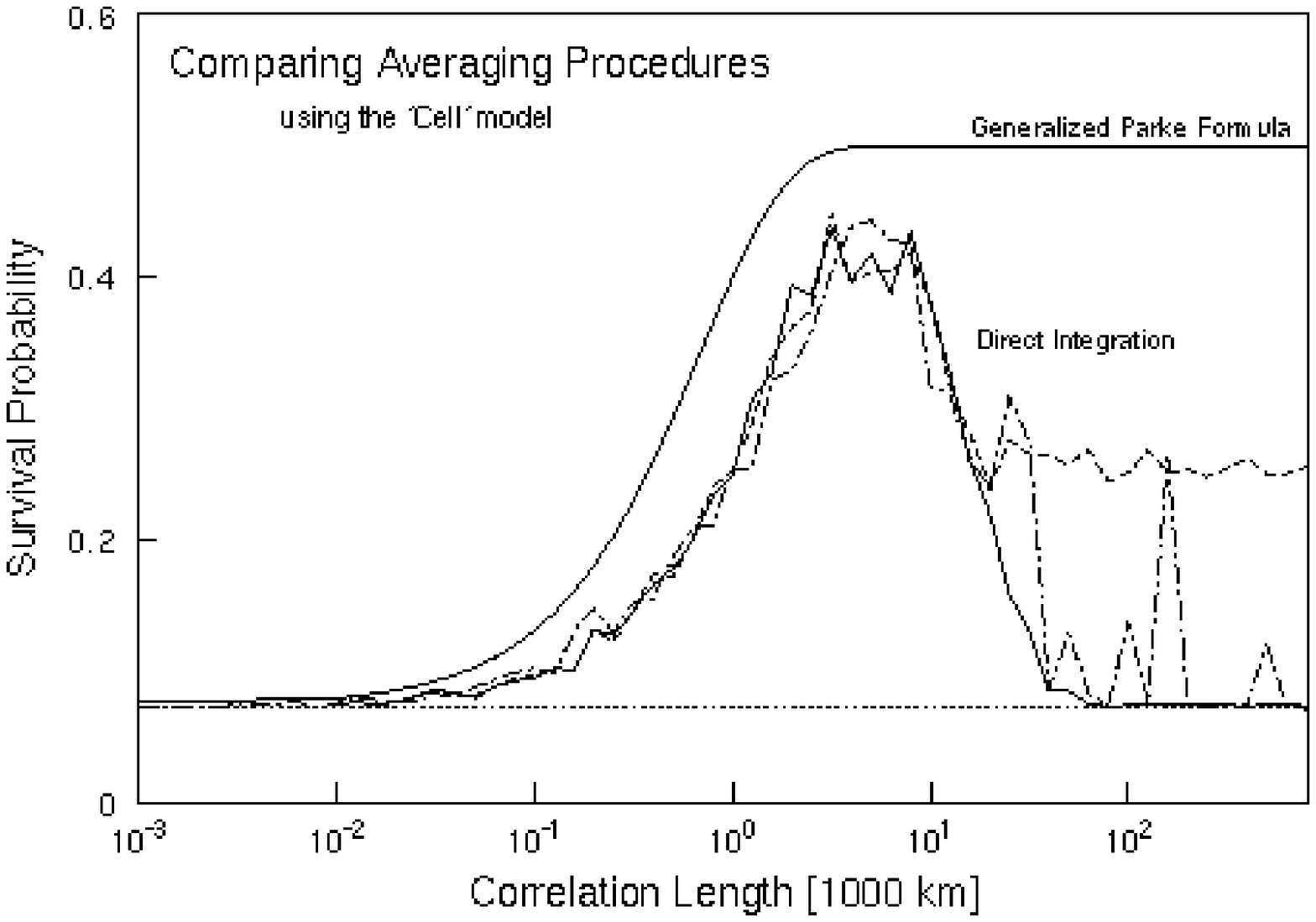}}

\begin{quote}
{\footnotesize               
{\bf Figure 2:} {\sl
A comparison of the generalized
Parke formula with the results of several numerical 
averages over a particular model -- the `Cell' 
model -- of solar fluctuations. The survival 
probability is plotted as a function of 
the correlation length, $\ell$, using the
representative values for solar neutrinos.
See ref.~\cite{petercliffdenis} for details.}}
\end{quote}

How badly does the Master equation describe
fluctuations which are larger than this? Very
badly indeed. Figure 2 shows a comparison of
the generalized Parke formula with a series of
direct numerical evaluations of neutrino evolution and 
the ensemble average, as reported in 
ref.~\cite{petercliffdenis}.
As may be seen from the figure, fluctuation effects
completely disappear once $\ell$ passes
appreciably above $\ell_{\rm crit}$. 

There is a qualitative argument for 
why fluctuations with very long correlation
lengths should not influence resonant
neutrino conversion. A density fluctuation
in the extreme limit $\ell \sim R_\odot$, 
simply corresponds to an overall shift of
the electron density throughout the entire
sun. This has the sole effect of simply
moving the position within the sun
where the the resonance occurs, but does not
alter the overall survival probability.  

In summary, the lessons to be 
drawn from the extant calculations of
neutrino oscillations within a fluctuating
medium are these:

\begin{enumerate}

\item 
Master equation methods describe solar
neutrinos well for sufficiently small correlation
lengths, but can dramatically overestimate the deviation
from the MSW prediction when correlation lengths
are large. For solar neutrinos, the dividing line 
between large and small correlation lengths is of 
order 3,000 km.

\item
Fluctuations can cause an appreciable deterioration
of the MSW resonance only if two conditions are
satisfied. First, fluctuation amplitudes cannot
be too small. Second, their correlation lengths 
should be neither too big nor too small
compared to the neutrino oscillation length. Most
importantly, both of these conditions must apply
{\it at the positions in the sun where the neutrino 
passes through the MSW resonance}.

\item
There is only one class of fluctuations 
in the sun which are known to exist with 
interestingly large amplitudes deep
within the solar radiative zone, where
the MSW resonance occurs. These are helioseismic
waves \cite{solwaves}, for which the correlation lengths
at resonance can be as large as $R_\odot/10 \sim
70,000$ km. As such they easily exceed the 
size of $\ell_{\rm crit}$, and so preclude a
description using only Master-equation techniques.

\end{enumerate}

\section{Helioseismic Waves}

We now summarize the results of detailed
simulations of neutrino oscillations in the presence
of helioseismic waves, as was originally reported
in ref.~\cite{petercliffdenis}. This reference 
explicitly computes seismic wave profiles within 
the sun in order to reliably
correlate their amplitude at the resonance point with
observations (which are made at the solar surface). It
then numerically evolves neutrinos through these
profiles to identify their effect on MSW oscillations.
No use is made of the short-correlation-length 
approximation.

There is a simple picture of what is going on. 
For any particular neutrino, microscopic fluctuations 
are irrelevant, and the perceived density profile
is static since the time taken for a neutrino 
to traverse the sun (several seconds) is much 
shorter than the typical period of a helioseismic 
wave (minutes to hours). The oscillations of
each neutrino are therefore well described by the ordinary
Parke formula, \ie\ eq.~(\ref{parke}) with $\lambda=1$. 
The dominant change in the survival probability
for each neutrino is then due to the differing
scale heights, $h$, (and hence differing jump
probabilities) which each neutrino sees. 
The average over the ensemble of density profiles 
seen by many neutrinos amounts to 
an average over $h$.

These calculations find that helioseismic waves
{\it have no detectable effect} on MSW oscillations,
even under optimistic assumptions concerning 
their amplitudes. This conclusion turns out to
rely in an important way on not using a 
Master-equation-based calculation, because 
those waves whose amplitudes can be large
enough to affect the neutrino resonance, also
have correlation lengths (wavelengths) which
are much larger than $\ell_{\rm crit}$.
More specifically:

\begin{enumerate}

\item 
Those helioseismic waves ($p$-waves) which
have already been observed at the solar surface
generally
{\it decrease} in amplitude as one moves into the sun,
with the result that their amplitude is too small 
in the MSW resonance region to produce observable
effects. 

\item
The main class of waves ($g$-waves) which are presently
undetected are also too small to affect neutrino
resonances if they carry energies which are
similar to the $p$-waves which have been seen. This
is so even though $g$-waves {\it grow} in amplitude 
with increasing solar depth.

\item 
A particular class of $g$-waves is believed
to be overstable, with runaway evolution 
once they are sufficiently excited. Although
the endpoint of the resulting runaway behaviour
remains controversial, it has been argued \cite{runaway}
that the energy in these modes could be as much
as $10^{10}$ times larger than what is in other modes. If
the generalized Parke equation were used to
propagate neutrinos through these waves, one
would predict detectable deviations from MSW
oscillations. Unfortunately, this prediction
is incorrect since the wavelengths of the overstable
modes are too large to permit a small-$\ell$
calculation.

\end{enumerate}

\section*{Acknowledgements}
The work described here is the result of very stimulating 
and pleasant collaborations with Denis Michaud and with 
Peter Bamert, to both of whom I am very grateful. Our 
principal funding comes from {\sl the Natural Sciences and 
Engineering Council of Canada}, with some additional funds 
being provided by {\sl les Fonds pour la Formation de Chercheurs 
et l'Aide \`a la R\'echerche du Qu\'ebec}.

% ------------------------------------------------------------------
% Journal abbreviations
% ------------------------------------------------------------------

\def\anp#1#2#3{{\it Ann.\ Phys. (NY)} {\bf #1} (19#2) #3}
\def\apj#1#2#3{{\it Ap.\ J.} {\bf #1}, (19#2) #3}
\def\arnps#1#2#3{{\it Ann.\ Rev.\ Nucl.\ Part.\ Sci.} {\bf #1}, (19#2) #3}
\def\cmp#1#2#3{{\it Comm.\ Math.\ Phys.} {\bf #1} (19#2) #3}
\def\ejp#1#2#3{{\it Eur.\ J.\ Phys.} {\bf #1} (19#2) #3}
\def\ijmp#1#2#3{{\it Int.\ J.\ Mod.\ Phys.} {\bf A#1} (19#2) #3}
\def\jetp#1#2#3{{\it JETP Lett.} {\bf #1} (19#2) #3}
\def\jetpl#1#2#3#4#5#6{{\it Pis'ma Zh.\ Eksp.\ Teor.\ Fiz.} {\bf #1} (19#2) #3
[{\it JETP Lett.} {\bf #4} (19#5) #6]}
\def\jpa#1#2#3{{\it J.\ Phys.} {\bf A#1} (19#2) #3}
\def\jpb#1#2#3{{\it J.\ Phys.} {\bf B#1} (19#2) #3}
\def\mpla#1#2#3{{\it Mod.\ Phys.\ Lett.} {\bf A#1}, (19#2) #3}
\def\nci#1#2#3{{\it Nuovo Cimento} {\bf #1} (19#2) #3}
\def\npb#1#2#3{{\it Nucl.\ Phys.} {\bf B#1} (19#2) #3}
\def\plb#1#2#3{{\it Phys.\ Lett.} {\bf #1B} (19#2) #3}
\def\pla#1#2#3{{\it Phys.\ Lett.} {\bf #1A} (19#2) #3}
\def\pra#1#2#3{{\it Phys.\ Rev.} {\bf A#1} (19#2) #3}
\def\prb#1#2#3{{\it Phys.\ Rev.} {\bf B#1} (19#2) #3}
\def\prc#1#2#3{{\it Phys.\ Rev.} {\bf C#1} (19#2) #3}
\def\prd#1#2#3{{\it Phys.\ Rev.} {\bf D#1} (19#2) #3}
\def\pr#1#2#3{{\it Phys.\ Rev.} {\bf #1} (19#2) #3}
\def\prep#1#2#3{{\it Phys.\ Rep.} {\bf #1} (19#2) #3}
\def\prl#1#2#3{{\it Phys.\ Rev.\ Lett.} {\bf #1} (19#2) #3}
\def\prs#1#2#3{{\it Proc.\ Roy.\ Soc.} {\bf #1} (19#2) #3}
\def\rmp#1#2#3{{\it Rev.\ Mod.\ Phys.} {\bf #1} (19#2) #3}
\def\sjnp#1#2#3#4#5#6{{\it Yad.\ Fiz.} {\bf #1} (19#2) #3
[{\it Sov.\ J.\ Nucl.\ Phys.} {\bf #4} (19#5) #6]}
\def\zpc#1#2#3{{\it Zeit.\ Phys.} {\bf C#1} (19#2) #3}

%\section*{References}

\end{document}